\documentclass[
preprint,
floats,
author-year,
]{revtex4}

\usepackage{amssymb}

\usepackage{graphicx}
\usepackage{color}		
\usepackage{epsfig}
\usepackage{ifthen}

\newlength{\goioswidth}
\begin{document}

\title[Safe and unsafe pathways on flutes]{ The kinetics and acoustics of
  fingering and note transitions on the flute} 

\author{André Almeida}
\affiliation{School of Physics, University of New South Wales, Sydney
  NSW 2052, Australia} 
\email{aalmeida@phys.unsw.edu.au} 

\author{Renee Chow}
\affiliation{School of Physics, University of New South Wales, Sydney
  NSW 2052, Australia} 
\email{renee.wei.yan.chow@gmail.com} 

\author{John Smith} 
\affiliation{School of Physics, University of New South Wales, Sydney
  NSW 2052, Australia} 
\email{john.smith@unsw.edu.au} 

\author{Joe Wolfe}
\affiliation{School of Physics, University of New South Wales, Sydney
  NSW 2052, Australia} 
\email{J.Wolfe@unsw.edu.au} 

\date{\today}
  

\begin{abstract}

  Motion of the keys was measured in a transverse flute while
  beginner, amateur and professional flutists played a range of
  exercises. The time taken for a key to open or close is typically 10
  ms when pushed by a finger or 16 ms when moved by a spring. Delays
  between the motion of the fingers were typically tens of ms, with
  longer delays as more fingers are involved. Because the opening and
  closing of keys will never be exactly simultaneous, transitions
  between notes that involve the movement of multiple fingers can
  occur via several possible pathways with different intermediate
  fingerings. A transition is classified as `safe' if it is possible
  to be slurred from the initial to final note with little perceptible
  change in pitch or volume. Some transitions are `unsafe' and
  possibly involve a transient change in pitch or a decrease in
  volume. In transitions with multiple fingers, players, on average,
  used safe transitions more frequently than unsafe
  transitions. Professionals exhibited smaller average delays between
  the motion of fingers than did amateurs.

\end{abstract}

\maketitle

\section{Introduction}
\label{sec:introduction}

In wind instruments, a single transition between two successive notes
often requires the coordinated movement of two or more fingers (for
simplicity, we shall refer to all digits, including thumbs, as
fingers). One of the reasons why players practise scales, arpeggi and
exercises is to learn to make smooth, rapid transitions between notes,
without undesired transients. For players, this motivation is
particularly important for slurred notes, where the transition
involves no interruption to the flow of air, and should ideally
produce no wrong notes and no detectable silence between the notes.

In practice, finger movements are neither instantaneous nor
simultaneous, and it takes a finite time to establish a new standing
wave in the instrument bore. Slurred transitions involving the motion
of only a single finger can produce transients that result from the
finite speed of the finger that pushes a key in one direction, or of
the spring that returns it to its rest position. For transitions
involving the motion of two or more fingers, there can be an
additional transient time due to the time differences between the
movements of each finger, which invites the question: how close to
simultaneous can flutist finger movements be, and are there preferred
finger orders in particular note transitions?

Although previous studies have monitored finger motion on the flute,
they have been concerned with the flute as a controller for electronic
instruments. The MIDI flute developed at IRCAM initially used optical
sensors, but the final version used Hall effect sensors with magnets
attached to the keys \citep{miranda_new_2006}. The ``virtually real
flute'' \citep{ystad_virtually_2001} used linear Hall effect sensors
and could detect the speed of key transitions. The hyper flute
\citep{palacio-quintin_hyper-flute_2003} employed a large number of
sensors, but only two keys had linear Hall effect sensors.
\textcite{palmer_movement_2007} used infra-red tracers attached to a
player's fingernails and recorded their motion with a video
camera. Although suitable for detecting the gestures of a player, this
approach does not provide sufficient resolution (in either space or
time) to monitor the detailed fingering behaviour occurring in note
transitions.

This paper reports explicit measurements of the times taken for keys
to open and to close under the action of fingers and springs, and
determines the key order and relative timing in transitions involving
multiple fingers. The flute was chosen partly because of the
similarity in size and construction of most of its keys, which means
that keys move with approximately similar speeds, and also that
similar sensors could be used for each. These sensors monitored the
position of each key using reflected, modulated infra-red radiation
and had the advantage that they did not alter the mass of keys nor
affect their motion. The flute has the further advantage that measured
acoustic impedance spectra are available for all standard fingerings
\citep{wolfe_acoustic_2001}, in addition to acoustical models of all
possible fingerings \citep{botros_virtual_2002,botros_virtual_2006}.

\section{Some background in flute acoustics}
\label{sec:some-backgr-flute}

In most woodwind instruments, the played note is determined in part by
the combination of open and closed holes in the side of its bore,
which is called a fingering. Each fingering produces a number of
resonances (corresponding to extrema in the input impedance), one or
more of which can be excited by a vibrating reed or air jet. On many
modern woodwinds there are more holes in the instrument than fingers
on two hands. Consequently some keys operate more than one tone hole,
often using a system of clutches, and some fingers are required to
operate more than one key.

The acoustical impedance spectrum of the flute for a particular
fingering can be predicted by acoustical models, and
important details of its behaviour can be deduced from this. The
'Virtual Flute' is an example of a web service using such an
acoustical model to predict the pitch, timbre and ease of playing
\citep{botros_virtual_2002,botros_virtual_2006}. This service
however does not yet give indications on the playing difficulties
associated with transitions between two different fingerings.

The embouchure of the flute is open to the air and so the instrument
plays approximately at the minima in the input impedance $Z(f)$ at the
embouchure. The player selects between possible minima by adjusting
the speed of the air jet \citep{coltman_jet_1976}, and consequently a
periodic vibration regime is established with fundamental frequency
close to that of a particular impedance minimum or resonance. Fine
tuning is achieved by rotating the instrument slightly, which has the
effect of changing the jet length, the occlusion of the embouchure
hole and thus the solid angle available for radiation, thereby
modifying the acoustical end effect. Changing from one fingering to
another usually changes most of the frequencies of the bore resonances
and consequently also the note
played. \textcite{de_la_cuadra_analysis_2005} discuss flute control
parameters in detail.

\subsection{The transitions between notes}
\label{sec:trans-betw-notes-1}

In some cases, no fingering changes are required: the player can
select among different resonances by adjusting the speed and length of
the jet at the embouchure. Thus the standard fingering for F4 is used
by players to play the notes F4 and F5, (and can also play C6, F6, A6
and C7).

Many of the transitions between two notes separated by one or two
semitones involve moving only a single finger. Provided that fingers
or springs move the key or keys sufficiently quickly, one would expect
no strong transients when slurring such transitions. Small transient
effects can always arise because of the fact that the strength of the
resonances in the bore of the flute are somewhat reduced when a key is
almost, but not completely closed (an example is given later in
Fig.~\ref{fig:FreqAmplKeyVStime}). 

Several mechanisms can produce undesirable transients in note
transitions. One of particular interest to us may occur when a slurred
transition involves the motion of two or more fingers. The speed of
moving keys is limited by the speed of fingers in one direction and of
the key springs that move them in the opposite direction. Further, the
acoustic effects produced by the motion of a key are not linear
functions of key displacement, so it is difficult to define
simultaneous motion, particularly for keys moving in opposite
directions. In practice, fingers will always move at slightly
different times and with different speeds (how different are these
times is one of the questions that we address). This means that,
between two notes that involve the motion of more than one finger,
there are several possible intermediate discrete key configurations,
as well as the continuous variations during key motion. Furthermore,
these different intermediate states may have quite different acoustic
properties, which are not necessarily intermediate between those of
the initial and final configurations.

\subsection{Safe and unsafe transitions}
\label{sec:safe-unsafe-trans}

\begin{figure}
 \centering
 \includegraphics[width=\goioswidth]{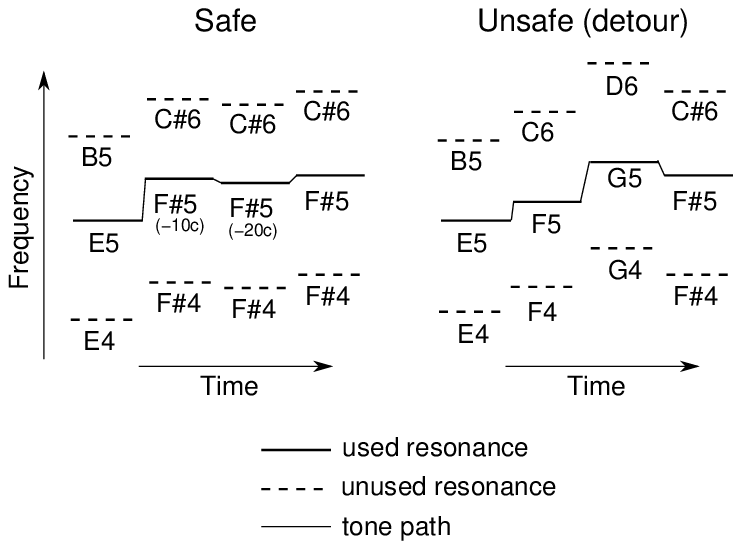}
 \includegraphics[width=\goioswidth]{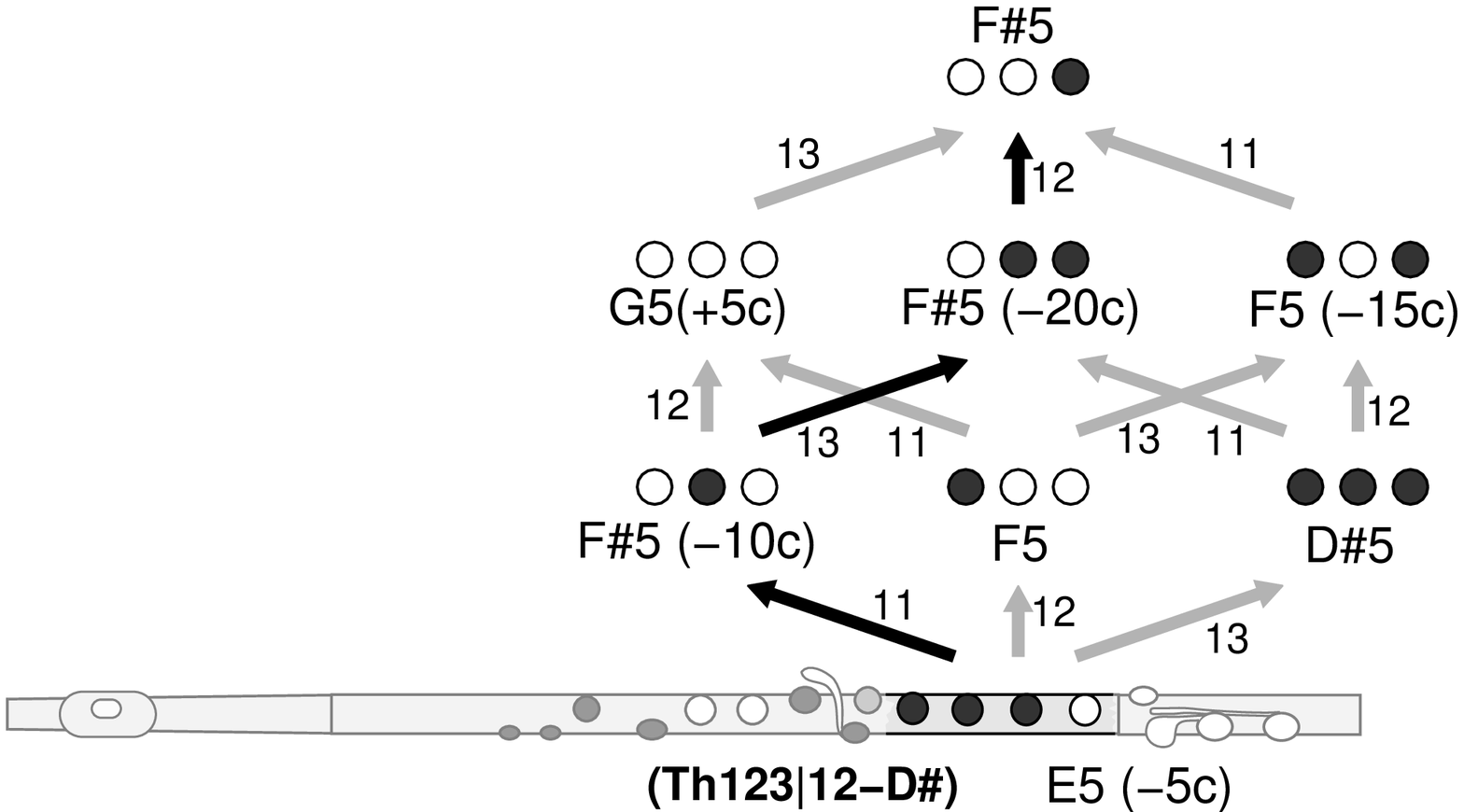}
 \caption{A schematic example of a safe transition and an unsafe
   (detour) transition from E5 to F$\sharp$5. In the safe transition,
   all intermediate fingerings produce notes with a pitch very close
   to that of the initial or final note. In the unsafe (detour)
   transition, the tone is not interrupted, but the transitory notes
   are not close to E5 or F$\sharp$5. The lower frame shows possible
   intermediate key states and transitions in that transition. The
   safe paths are shown with dark arrows, and labelled with number of
   the key that moves. White circles indicate open tone holes, black
   indicates holes closed by a key directly under the finger, and grey
   shows those closed indirectly by the mechanism. The legends show a
   common notation for these fingerings. }
 \label{fig:unsafeA}
 \label{fig:statesE5F5}
\end{figure}

\begin{figure}
 \centering
 \includegraphics[width=\goioswidth]{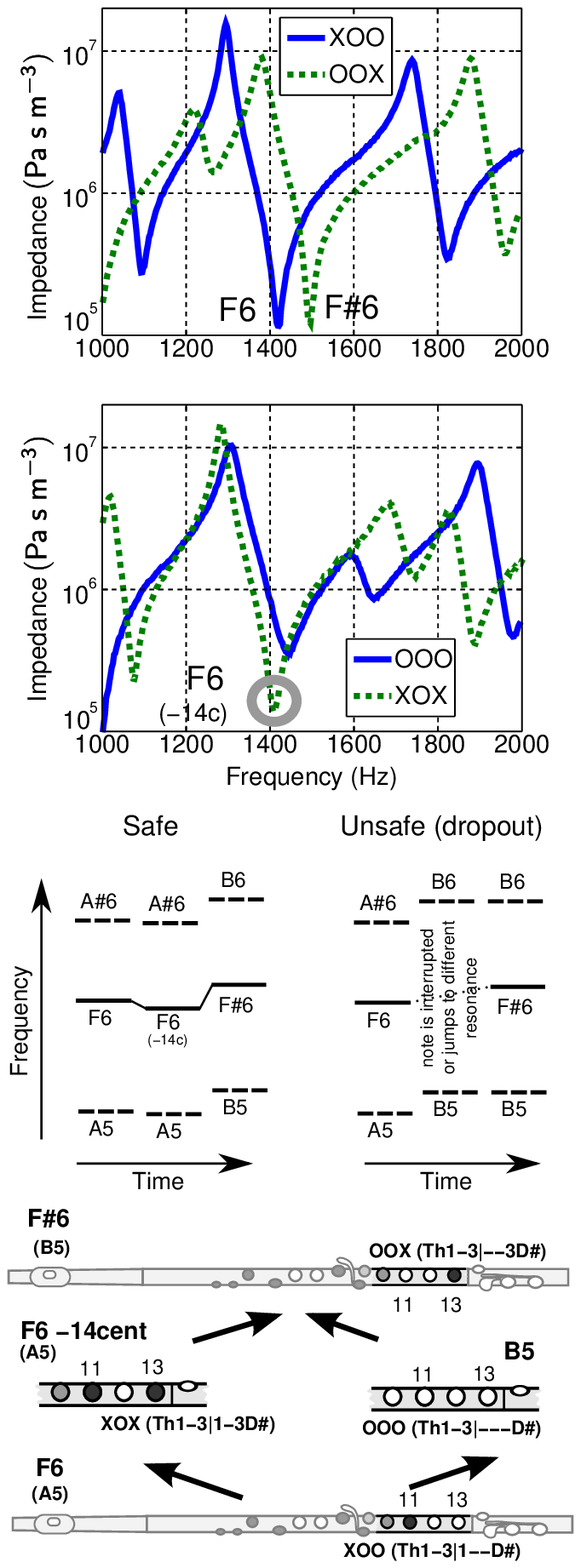}
 \caption{Schematics illustrating a safe and an unsafe (dropout)
   transition from F6 to F$\sharp$6. The top graph shows the impedance
   spectra, showing the minima that play F6 and F$\sharp$6, and the
   second graph the impedance spectra for the possible intermediate
   fingering states.  The possible transition paths are shown in the
   schematic graph of frequency vs time. The fingerings involved are
   shown in the lowest schematic. The keys controlled by the right
   hand are emphasised and the rest shown pale.   }
\label{fig:unsafeB}
\end{figure}

We divide intermediate fingerings into three categories:

\begin{description}
\item[SAFE.] If the minima for $Z(f)$ lie at frequencies close to (or
  intermediate between) those of the initial and final states, and
  have similar magnitudes, then a steady oscillation of the jet can be
  maintained during a slurred transition. A transition that passes
  transiently through such a fingering can be called a safe transition
  and is illustrated schematically in Fig.~\ref{fig:unsafeA}. This
  result is discussed in more detail in section
  \ref{sec:two-fing-trans}.
\item[UNSAFE (detour).] If one of the intermediate states exhibits a
  minimum in $Z(f)$ at a frequency unrelated to both notes in the
  desired transition, it may cause an undesired note to sound briefly
  during the transition -- see Fig.~\ref{fig:unsafeA}.
\item[UNSAFE (dropout).] If there are no deep minima for $Z(f)$ at
  frequencies close to those of the initial and final states, jet
  oscillation may not be maintained during a slurred transition,
  because the jet length and speed are inappropriate for the frequency
  of the nearest minimum. In this case, the intensity of the tone will
  decrease substantially. Fig.~\ref{fig:unsafeB} shows an example of a
  note transition for which one of the intermediate fingerings has a
  weak resonance.

\end{description}

We could also define a transition as `unsafe (trapped)' when the
second fingering has, as well as the resonance usually used, a strong
resonance at a frequency close to that of the first note and so may
`trap' the jet. \textcite{botros_virtual_2003} gives examples of this
situation.

In some cases, such as the C6 to D6 transition (also analysed later),
there is no safe intermediate fingering so, unless fingers move nearly
simultaneously, audible transients are expected. Of course, even for
our definition of safe transitions, transients are expected in the
flute sound: it takes time for a wave to travel down the bore, to
reflect at an open tone hole and to return, and several such
reflections may be required to establish a standing wave with a new
frequency. Finally, it should be remembered that some transients are
an important part of the timbre of wind instruments and may be
expected by musicians and listeners.

\section{Materials and methods}
\label{sec:materials-methods}

\subsection{Monitoring the key positions }
\label{sec:monit-key-posit}

An optical method was chosen because, unlike magnetic systems, there
was no need to attach magnets or other material to the keys, and thus
alter their mass or feeling under the fingers. Mechanical contacts
suffer reliability problems and exhibit bounce and/or hysteresis.

\begin{figure}
  \centering
  \includegraphics[width=\goioswidth]{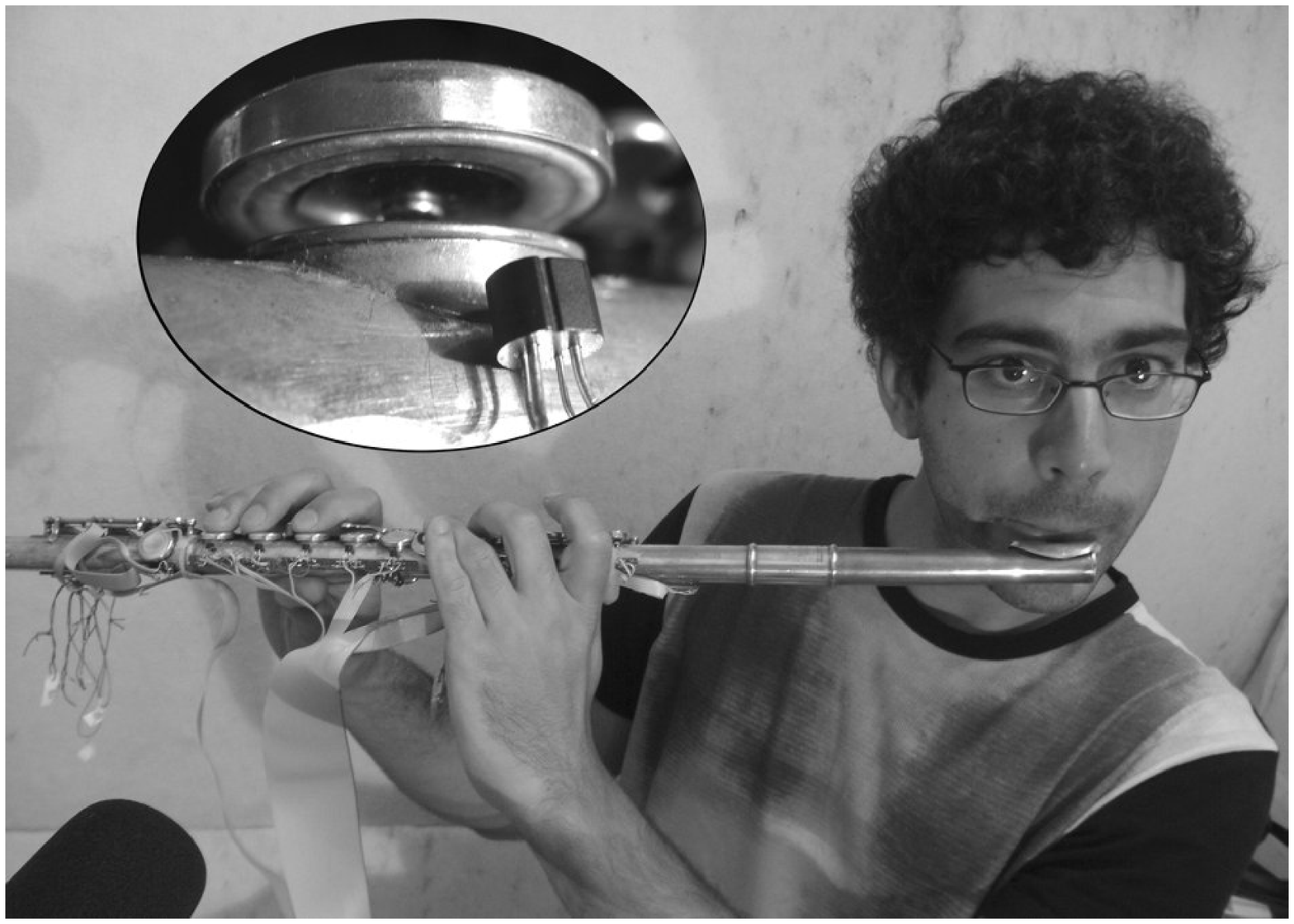}
  \caption{An author (AA) demonstrates the modified flute used for
    this work. The 50 wire IDC cable that connects the flute to the
    sensor electronics is visible below the flute. The microphone can
    be seen in the lower left corner. The inset shows how a
    photosensor is mounted below a key.}
  \label{fig:flutedemo}
\end{figure}

A reflective photosensor was glued below the edge of each key so that
the intensity of light reflected from the edge of the key increased
monotonically as the key was closed (see
Fig.~\ref{fig:flutedemo}). The chosen sensors (Kodenshi SG-2BC) were
small (4 mm diameter), low mass (160 mg), and combined an infrared
light emitting diode (LED) with a high-sensitivity phototransistor
(peak sensitivity at 940 nm). Instead, the sensor signal was
distinguished from the background illumination by modulating the
output of the LED in the sensor with a 5 kHz sine wave. The
phototransistor in the sensor was connected as an emitter follower
with an RC stage to filter out DC variations due to changes in
lighting conditions. Because the background illumination and degree of
shading can vary for each experiment, the DC bias was adjustable to
provide adequate dynamic range for the 5 kHz signal without
clipping. This was set using a separate 8 element bar LED display for
each channel. This procedure removes most, but not all, of the
dependence on background illumination: a small component remains
because of non-linearities. The sensor signals from 16 keys and the
sound were recorded on a computer using two MOTU 828 firewire audio
interfaces (24 bit at 44.1kHz).  Because the same hardware was used to
sample both the sound and the output of the key sensors, we can be
certain that any delays in sampling due to latencies and buffering
will be identical and consequently will cancel exactly when timing
differences are calculated.

\begin{figure}
 \centering
 \includegraphics[width=\goioswidth]{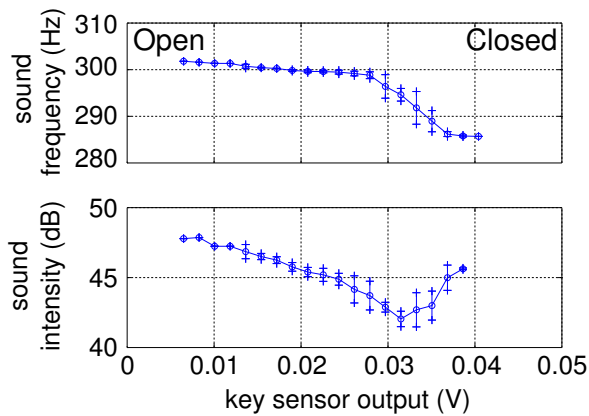}
 \caption{The variation of sound frequency and intensity produced by a
   flute when a key is slowly closed. The air jet was provided by a
   blowing machine. The saturation of the frequency (top graphic) with
   increasing sensor value at the right hand side of the curve is the
   result of continuing compression of the key pad with no further
   acoustic effect.  The bottom graphic shows how the intensity is
   reduced when the pad is almost but not completely closed. Error
   bars indicate the standard deviation for 6 different trials.}
 \label{fig:FreqAmplKeyVStime}
\end{figure}

The sensor output as a function of position was measured in
experiments in which the sensor output was recorded while a key was
slowly closed using a micrometer screw. These showed that the
amplitude of the modulated output from the sensor was a monotonic, but
nonlinear function of key position. In a further series of
experiments, the flute was played by a blowing machine while a key was
slowly closed by the micrometer. The frequency and intensity are
plotted as a function of sensor output in
Fig.~\ref{fig:FreqAmplKeyVStime}. The shape of this plot is explained
by the compressibility of the key pad. Once the pad makes contact with
the rim of the tone hole chimney, the playing frequency reaches its
lowest value and remains unchanged as the pad is compressed while the
key is further depressed. (The sound level is high when the key is
fully open and also when the key is closed with the pad
compressed. The sound level is lower, however, when the hole is almost
closed by the uncompressed pad, which is presumably a consequence of
leaks between the pad and the rim, this can be seen in
Fig.~\ref{fig:FreqAmplKeyVStimeMulti})

Measurements such as that shown in Fig.~\ref{fig:FreqAmplKeyVStime}
cannot be simply used to calibrate measurements made with players
rather than blowing machines. The main reason is that, for some
players, the tip of the finger occasionally overhangs the key and
contributes a small component to the reflection measured by the
sensor. 

\subsection{Recording and analysing the sound}
\label{sec:record-analys-sound}

The sound produced by the flute was recorded using a Rhode NT3
microphone placed on a fixed stand and digitized using one input of
the MOTU audio interfaces. The midpoint of the flute was approximately
50 cm away from the microphone. The frequency was calculated from the
recorded sound file using Praat sound analysis software
\citep{boersma_accurate_1993}, using autocorrelation with an analysis
window of 15 ms. The time resolution for frequency transitions was
estimated to be 8 ms by studying a semitone discontinuity in a
sinusoidal signal (i.e. between 1000 Hz and 1059 Hz). The intensity
and sound level were also extracted using Praat.

\subsection{Characterisation of open/closed states and note
transitions }
\label{sec:char-opencl-stat}

One of the purposes of this study concerns the relative timing of the
open/closed transitions, so it is necessary to relate a defined value
of sensor output to the effective opening/closing of the associated
key. Most of the variation in sound frequency occurs close to the
point of key closure, so the saturation point in
Fig.~\ref{fig:FreqAmplKeyVStime} was considered as a possible choice.
In practice, because of the variations described earlier, this value
would be somewhat different for each flutist, key and level of
background illumination. Instead, guided by curves such as those shown
in Fig.~\ref{fig:FreqAmplKeyVStime}, we set the reference value for a
key transition between 70\% and 85\% of the total variation in sensor
output, the exact value depending on the key and situation (see
Fig.~\ref{fig:FreqAmplKeyVStime}).

\begin{figure}
  \centering
  \includegraphics[width=\goioswidth]{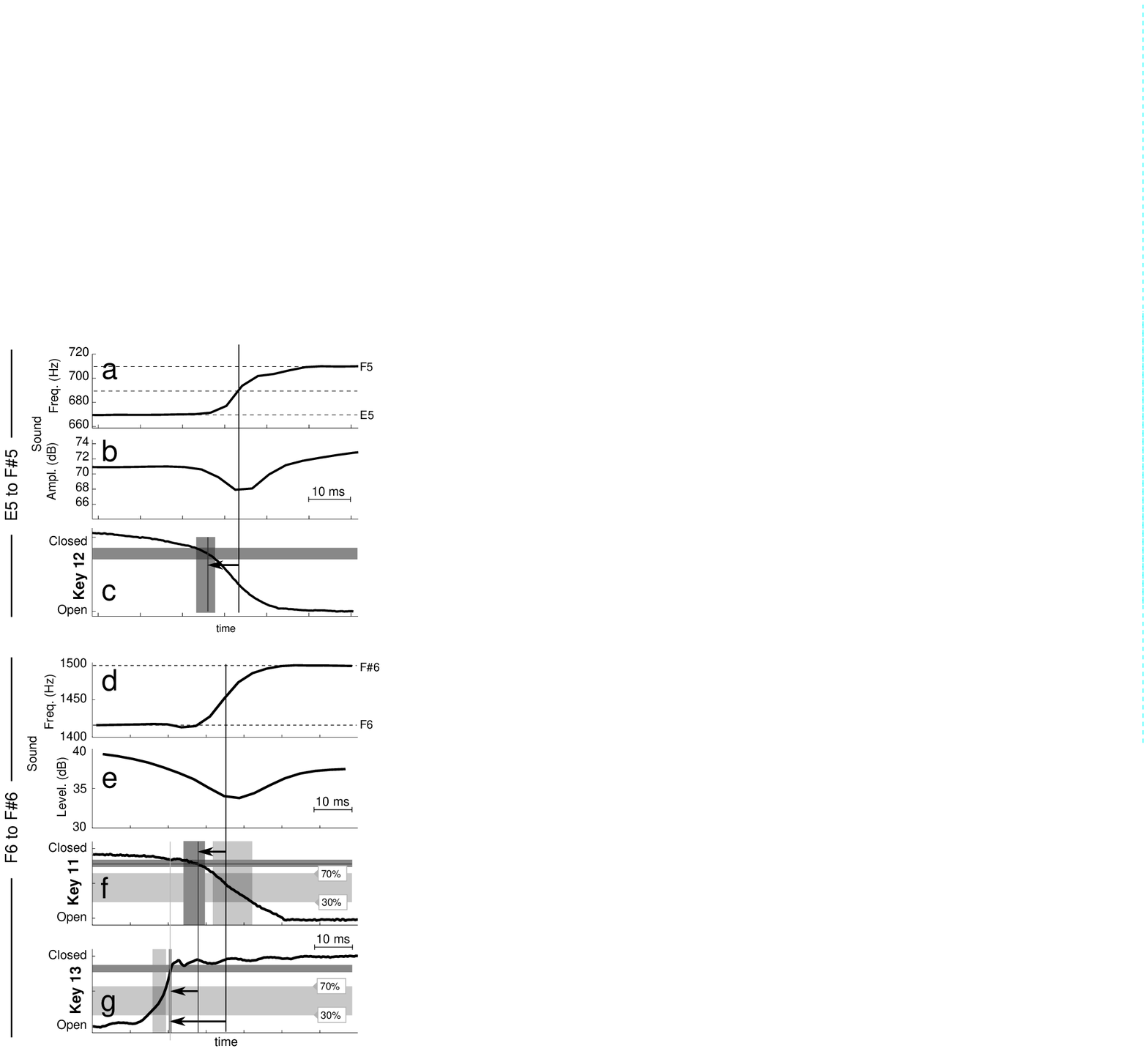}
  \caption{Examples of parameter extraction from the measured sound
    and key sensor output. Typical measurements of the frequency
    (\textbf{a}), sound level (\textbf{b}) and key sensor output
    (\textbf{c}) are extracted for a single finger transition from E5
    to F5, corresponding to the movement of a single key.  The
    horizontal dark shaded band shows the uncertainty in the key
    transition value, and consequently the vertical shaded band shows
    the uncertainty in the time of transition. The horizontal arrow
    shows the offset between effective key opening and the midpoint of
    the frequency transition. A two finger transition from F6 to
    F$\sharp$6 is shown in parts \textbf{d} to \textbf{g}.  Their
    sensor values are shown in \textbf{f} and \textbf{g}.  The dark
    shaded bands again show the uncertainty in the key transition
    value and time. The pale shading shows the time between 30\% and
    70\% of the key sensor value, which we discuss in the text. Again,
    the arrows show the interval between note transition and effective
    key opening/closing. The difference between these arrows gives the
    delay between the two keys, here about 8 ms.}
\label{fig:FreqAmplKeyVStimeMulti}
\end{figure}

Determining the duration of effective key opening and closing is also
complicated by the saturation of sensor output described above. After
examination of a range of traces under different conditions, we chose
to measure the time taken between sensor signals of 30\% and 70\% of
the maximum range of the sensor output. This rate of change was then
multiplied by a factor of 100/40 to produce a measurement of the
effective key closing or opening time. Examples are shown in
Fig.~\ref{fig:FreqAmplKeyVStimeMulti}.

An automated software routine was used to detect and characterise the
key movements in each recording. First, it detected each time the
output of a key sensor passed through a value midway between
neighbouring fully closed and fully open states. These then served as
initial starting points to find the nearest times when a key was
effectively open or closed. These allowed calculation of the duration
of each open/closed or closed/open transition. We estimated an
uncertainty in each individual measurement by determining how long it
took each key sensor output to vary by $\pm$5\% around the effective
open or closing value. This value, was on average 1.4$\pm$1.4 ms
(n=2069) for a closing key and 4.1$\pm$2.6 ms (n=1639) for an opening
key.

A single key may be associated with several different note
transitions, so the key movements, detected as described above, need
to be associated with the note transitions of interest. In order to
find note transitions automatically, the detected pitch was quantised
to the set of notes used in the exercise. These quantised data were
smoothed using a filter which calculates the median value within a
window of 50 ms so that only sufficiently long values corresponding to
note transitions are detected. For each note transition thus detected,
the corresponding nearest transition reference times for key motion
are found. The pitch and key position detection are shown for two
particular exercises in Fig.~\ref{fig:FreqAmplKeyVStimeMulti}.

The following parameters were calculated for each key transition
associated with a given note transition: the effective duration of the
key transition, the offset in time between the key transition and the
pitch transition (see Fig.~\ref{fig:FreqAmplKeyVStimeMulti}) and an
estimate of the uncertainty in the key closing or opening time.

\subsection{Subjects}
\label{sec:subjects}

The 11 players participating in this experiment were divided into 3
categories according to experience. Professionals (players P1 to P7)
are those with more than 8 years experience, and for whom flute
playing is a significant part of their current professional life,
either as performers or teachers. Amateurs (A1--A3) have
between 3 and 8 years of flute playing experience, and regularly play
the flute as a non-professional activity. Beginners (B1) have less
than 3 years experience.

\subsection{Experimental protocol}
\label{sec:exper-prot}

These experiments were conducted in a room with a low reverberance
that was significantly isolated from external noise. All measurements
were performed on a specific flute from the laboratory, a C-foot Pearl
flute fitted with a sensor near each key. The flute is a plateau or
closed key model, i.e. the keys do not have holes that must be covered
by the fingers and it has a split E mechanism, which means that there
is only one hole functioning as a register hole in the fingering for
E6. Players could use their own head joint if desired.

A typical session took about 75 minutes. Each subject was asked to
warm up freely for about 15 minutes, in order to become accustomed to
the experimental flute, the change in balance and some awkwardness
caused by the cables. They also used this time to rehearse the
particular exercises in the experimental protocol. The musical
exercises, written in standard musical notation, were delivered to the
subjects approximately one week before the recording session. Some
players did not complete all exercises.

\subsection{Experimental exercises}
\label{sec:exper-exerc}

The players were recorded performing a selection of note transitions,
scales, arpeggi and musical excerpts. Except for the musical excerpts,
each written exercise was performed at least twice, at two different
tempi: players were asked to play ``fast'' (explained to them thus: as
fast as possible while still feeling comfortable and sure that all the
notes in the exercise were present) and ``slow'' (in a slow tempo but
still comfortable to perform the exercise once in a single breath). In
the case of the fast performance, the musician was asked to repeat the
exercise as many times as possible (but still comfortable) in a single
breath.

\section{Results and discussion}
\label{sec:results-discussion}

The times taken for each key to be effectively depressed or released
(i.e. to make the relevent key open or close depending upon its
mechanism) are shown in Table \ref{tab:dur}. In all cases, pressing
times are quicker, perhaps because the finger can transfer momentum
gained in a motion that may begin before contact with the key, whereas
a released key has to be accelerated from rest by a spring. The large
variation among the durations for finger activated motion may include
the variations in strength and speed of the fingers in the somewhat
different positions in which they are used. There is rather less
variation among the mechanical release times. However, some keys
differ noticeably from the others.  Large variation in the latter
involves the key mechanism: some keys move alone, others in groups of
two or three, because of the clutches that couple their motion. In
particular, the left thumb key and the right little finger (D$\sharp$
key) have stiffer springs, so their release movement is significantly
faster ($p<0.001$) than for other keys. Overall, slow tempi produce
significantly slower ($p=0.03$) key press times.

\begin{table}
  \centering
  \begin{tabular}{rlcc}
    Key & Finger & Press time (ms) & Release time (ms)\\
    3 & L Index & 11.3$\pm$4.8 (100) & 15.9$\pm$4.9 (100)\\
    4 & L Thumb & 8.7$\pm$1.1 (204) & 9.2$\pm$2.1 (205)\\
    6 & L Medium & 15.2$\pm$6.0 (101) & 16.9$\pm$3.7 (304)\\
    7 & L Ring & 17.0$\pm$9.5 (273) & 22.2$\pm$9.4 (480)\\
    11 & R Index & 11.2$\pm$5.1 (160) & 15.6$\pm$3.3 (185)\\
    12 & R Medium &  8.9$\pm$11.1 (160) & 15.3$\pm$3.1 (181)\\
    13 & R Ring &  8.3$\pm$2.5 (532) & 16.7$\pm$9.4 (524)\\
    14 & R Little &  12.1$\pm$2.6 (179) & 12.9$\pm$9.3 (172)\\
  \end{tabular}

  \caption{Durations (average $\pm$ standard deviation and sample size in brackets) of different key movements}
  \label{tab:dur}
\end{table}

\subsection{Single finger transitions}
\label{sec:single-fing-trans}

When only one key is involved in a note transition, the pitch change
is a direct consequence of the motion of that key. As explained above,
the transition from one note to another is not discrete. The frequency
of the impedance minimum corresponding to the fundamental of one note
is shifted gradually as the opening cross-section of the hole is
increased. A relatively small range of key positions, near the fully
closed limit is associated with most of the change in pitch
(Fig.~\ref{fig:FreqAmplKeyVStime}): variations in position near the
fully open state have much less effect.  The delay between detected
key motion and frequency change was 1.9$\pm$3.8 ms (n=1303), which is
less than the uncertainties in the measurements: the uncertainty in
frequency change is several ms and the experimental uncertainty in key
motion a few ms (Fig.~\ref{fig:FreqAmplKeyVStimeMulti}). (The time for
radiated sound from the flute to reach the microphone was only about 1
or 2 ms.)

\subsection{Two finger transitions}
\label{sec:two-fing-trans}

\begin{figure}
  \centering
  \includegraphics[width=\goioswidth]{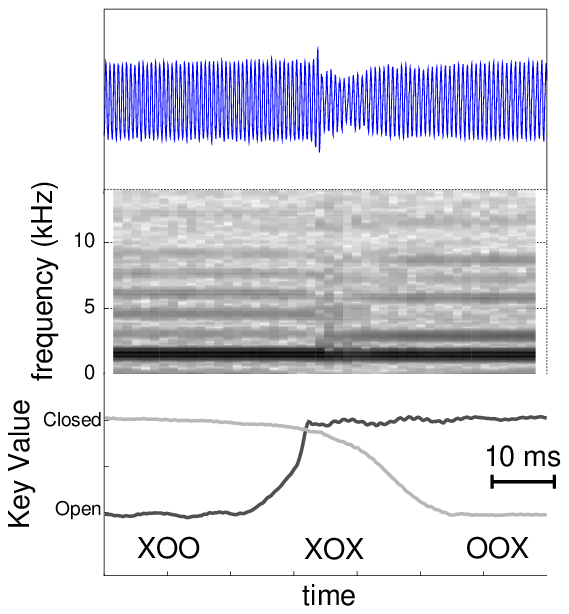}
  \includegraphics[width=\goioswidth]{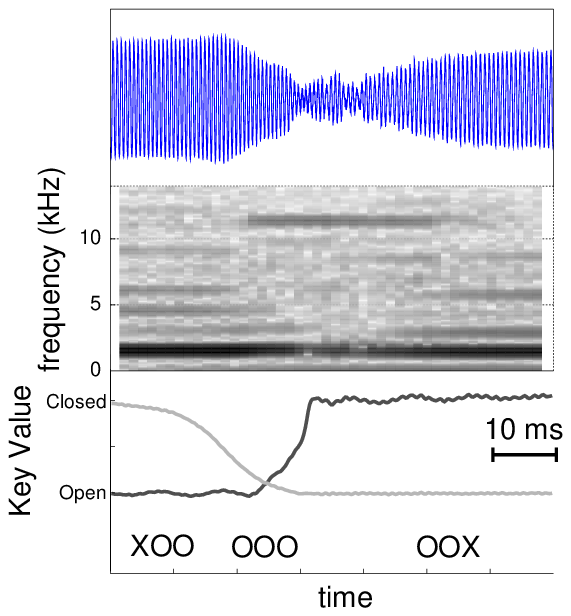}
  \caption{Oscillograms and spectrograms of the sound produced by the
    same player in two examples taken from the same trial for  F6
    to F$\sharp$6 transitions (nominally 1397 to 1480 Hz), with key
    sensor signals shown below. Spectra were taken with windows of 1024
    samples (23 ms), separated by 256 samples (6 ms), gray levels are
    proportional to the logarithm of the amplitude in each frequency
    bin. The example shown on top with both keys closed during the
    transient is a safe transition (see text), and that on the bottom
    with both keys open during the transient is an unsafe
    (dropout). The former shows a brief and relatively continuous
    transient. In the latter, the harmonic components of the sound are
    interrupted for tens of ms. During this transient, a larger band
    component appears at about 12 kHz, unrelated to the resonances of
    the bore, corresponding to the edge tone produced in the absence
    of a suitable resonance in the unsafe transition.}
  \label{fig:oscillograms}
\end{figure}

When two keys are involved in a note transition, there are two
possible intermediate key configurations due to the non-simultaneous
movement of the fingers. Examples involving the index and ring fingers
of the right hand moving in opposite directions are the transitions F4
to F$\sharp$4, F5 to F$\sharp$5 and F6 to F$\sharp$6, but only
F6/F$\sharp$6 involves an unsafe intermediate. Using X to indicate
depressed and O unpressed, and only indicating the first three fingers
of the right hand, this transition goes from XOO to OOX, with the
possible intermediates being XOX and OOO, as shown in
Fig.~\ref{fig:unsafeB}. The fingering with both keys depressed (XOX)
plays a note 14 cents flatter than F6. The fingering with neither key
depressed (OOO) is more complicated. If the speed of the air jet is
well adjusted to play F6 or F$\sharp$6, then this fingering does not
play a clear note (see Fig.~\ref{fig:oscillograms}). If the speed of
the air jet is somewhat slower than would normally be used to play
these notes, then it will play a note near B5. So, apart from the
ideal, unrealisable, `simultaneous' finger movement, there can only be
one safe path for the transition XOO to OOX and that goes via the
fingering XOX (in which both keys are briefly depressed): the slight
perturbation in pitch cannot be detected in a rapid transition.

We have also sought to compare intermediate states used during
different exercises involving the same transition. Players were asked
to rapidly alternate between the two notes as well as play it in the
context of a scale. The exercise of rapidly alternating between XOO
and OOX fingerings is an artificial exercise.  For such rapid
altenations, players often use special trill fingerings, in which
intonation and/or timbre are sacrificed in return for ease of rapid
performance. To perform this trill, a flutist would normally alternate
the XOX and OOX fingerings, i.e. transform it into a single finger
transition using the ring finger only. Thus the exercise is one that
flutists will not have rehearsed before this study. By contrast, the
same key transition in the context of a scale (here the B$\flat$ minor
scale), is one which experienced flutists will have practised over
years.

\begin{figure}
  \centering
  \includegraphics[width=\goioswidth]{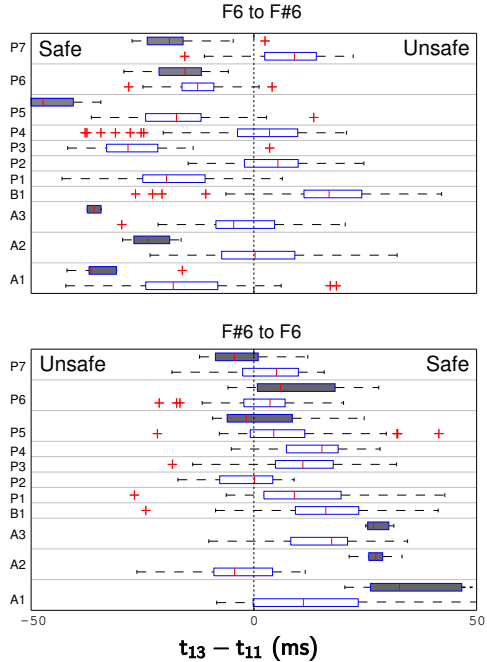}
  \caption{`Box plot' representation of time differences between keys
    13 and 11 for transitions between F6 and F$\sharp$6 in
    alternations (hollow rectangles) and scale exercises (filled
    rectangles) for different players. Rectangles represent the range
    between the first and third quartiles with a central line
    representing the median. Error bars extend to the range of data
    points not considered as outliers. Outliers are represented as
    crosses. }
  \label{fig:histograms}
\end{figure}

Considerable differences were found between slow and fast trials (data
not shown).  The results for all players are presented in
Fig.~\ref{fig:histograms}.

\label{sec:statistics}

Fig.~\ref{fig:histograms} shows that the descending transition
(F$\sharp$6 to F6) has a relatively consistent behaviour. For note
alternations, professional musicians used a safe finger order 72\% of
the times (that is transiting through the XOX state where both keys
are depressed).  Although they sometimes (48\%) use the unsafe finger
order in the scale context, $t_{13} - t_{11}$ was in average
3.8$\pm$9.6 ms (n=210), so that this transition is close to
simultaneous.

In the ascending case (F6 to F$\sharp$6) of the alternation exercise
professionals used safe transitions in 57\% of cases, although the
behaviour was less homogeneous among players (p<0.001 in the
F6/F$\sharp$6; p=0.02 in F$\sharp$6/F6), but in the scale context
musicians used safe transitions 97\% of the time ($\langle t_{13} -
t_{11}\rangle$ = -26$\pm$17 ms (n=33)).

Although the transition from F4 to F$\sharp$4 uses the same fingerings
as F6 to F$\sharp$6, in this case all transition pathways are
safe. Interestingly, most of the professionals tend to use the finger
order which would be safe for F6 to F$\sharp$6 with similar time
differences between keys (p=0.02 for F/F$\sharp$ and p=0.13 for
F$\sharp$/F), even though there is no unsafe intermediate for F4 to
F$\sharp$4. For one professional (P5) and most amateurs, the time
differences did change significantly but with no consistent direction
(either becoming more or less simultaneous). The finger order remained
the same as for the other subjects.

\subsection{Three finger transitions }
\label{sec:trans-involv-more}

The example we will use is the E5 to F$\sharp$5 transition, which
involves three different keys (11, 12 and 13), moved respectively by
the index, middle and ring fingers of the right hand, with the little
finger remaining depressed throughout. Using the notation described
above and neglecting other keys, E5 is played using XXO
(\mbox{Th123|12-D$\sharp$}) and F$\sharp$5 is played OOX
(\mbox{Th123|-$\,$-3D$\sharp$}). Thus the fingers are lifted from the
keys 11 (index) and 12 (middle) and key 13 (ring) is depressed.

Discounting the idealised simultaneous movement of the fingers, there
are six possible pathways involving discrete transients. From an
acoustical point of view, only one of these is safe: key 11 moves
first, then key 13 then key 12 (i.e. XXO, OXO, OXX, OOX). This path is
safe because OXO and OXX both play slightly flat versions of
F$\sharp$5. Conversely, when descending from F$\sharp$5 to E5, the
only safe transition is 12, 13 and then 11 (OOX, OXX, OXO, XXO). Any
other path involves a fingering that produces a note near G5 (OOO), F5
(XOX or XOO) or D5 (XXX), which may or may not be noticeable depending
on the duration of the intermediate. The paths are shown in
Fig.~\ref{fig:statesE5F5}.

\begin{figure}
  \centering
  \includegraphics[width=\goioswidth]{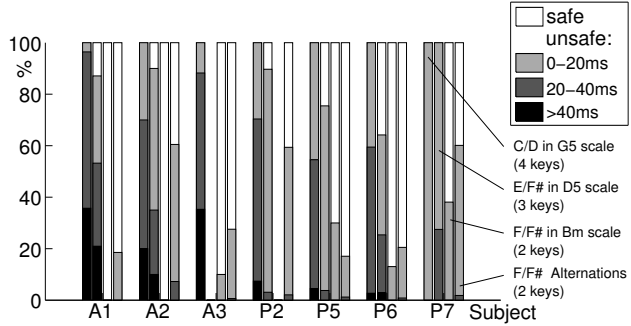}
  \caption{The percentage of safe and unsafe note transitions. The
    different shaded sections of each bar show the data for three
    amateurs (left) and four professionals (right). For each player,
    the group of four vertical bars represents transitions involving
    contrary motion of fingers: C6-D6 in a G scale (four fingers),
    E5-F$\sharp$5 in a D scale (three fingers), and F6-F$\sharp$6 in a
    B minor scale and also in an alternation exercise (two fingers).
  }
  \label{fig:UnsafetyOveral}
\end{figure}

The results for all flutists for this transition are summarised in
Fig.~\ref{fig:UnsafetyOveral}. For this transition, most professionals
are nearly safe, using pathways that are unsafe for only 20 ms. The
durations in the intermediate states vary substantially among
players. Even though this context (a D major scale) is one that
flutists would have practised many times, the delay times are
substantially longer here than for the two finger, contrary motion
example shown above, which uses two of the same fingers. From
examining the average and standard deviations of the inter-key time
difference, we also observe that some of the subjects (P2 and P5) have
significantly smaller time differences (data not shown). To summarise,
professionals spend an average of 13 $\pm$ 9 ms (n=225) in unsafe
transitions whereas amateurs spend 25 $\pm$ 16 ms (n=88), which are
significantly different ($p<0.001$). 

\subsection{Four finger transitions}
\label{sec:4-finger-transition}

The three long fingers and the thumb of the left hand move in the
transition from C6 (OXOO, or \mbox{1-$\,$-|-$\,$-$\,$-D$\sharp$}) to
D6 (XOOO or \mbox{Th-23|-$\,$-$\,$-D$\sharp$}), with the index finger
releasing a key and the others depressing keys. Here, there is no
completely safe path of transient fingerings, because the intermediate
states involving a change in position of any two fingers all sound a
note different from C6 or D6. There are partially safe paths, for
example, starting with C6 and moving first either the middle or ring
finger still produces a note very close to C6.

Average times spent in unsafe transitions, measured in the context of
a G major scale, are shown in Fig.~\ref{fig:UnsafetyOveral}. As with
the previous example, most flutists exhibited some preferred paths,
but they are not consistent among players, and sometimes the same
player may use different finger orders while playing quickly or
slowly.

To summarise, all professional players spend less time in unsafe
configurations than do the amateurs: unsafe intermediate states last
for an average of 34 $\pm$ 14 ms (n=65) for amateurs, 21 $\pm$ 11 ms
(n=133) for professionals. The difference is significant
($p<0.001$). Thus, for both professionals and amateurs, the time spent
in unsafe configurations is larger when four fingers rather than three
are involved.

\subsection{Summary of multi-finger transitions}
\label{sec:summary-multi-finger}

Typical values of delays between fingers are about 10 to 20 ms for
transitions that involve the motion of two or more fingers and
significantly longer for amateurs than for professionals. In
multi-finger transitions, the most frequent finger orders are often
those that avoid or minimise the use of transient fingerings that are
unsafe, as defined above. This is particularly true for transitions
involving two fingers, but less evident in more complicated
transitions.

We can compare the finger motion delays for the four, three and
two-finger changes discussed above. (For one finger, the delay by
definition is zero, as it does not include the time for key motion.)
This is shown in Fig.~\ref{fig:UnsafetyOveral}, which summarises the
results obtained in the examples studied in this article.

These results can be related to similar studies on repetitive tapping
movements in non-musical exercises \citep{aoki_differences_2003}. In
these, single finger movements show a variability in inter-tap
intervals of about 30 ms, increasing to 60 ms in the most agile
fingers when two fingers are involved. When ring and little fingers
are involved, this value is increased to 120 ms. These high values
suggest that the inaccuracies in multi-tap intervals are mostly due to
the duration of the finger motion rather than to synchronisation
between the motion of two fingers, but no references were found for
measurements of these values.

Some informal tests on our subjects show that when no sound output is
involved the standard deviation in delays between keys can increase
from approx.~20 to 40 ms, independently of the proficiency. These values
suggest that in musicians the sound feedback improves the performance
of the gesture.

Finally, delays between fingers are unimportant if their effect on
sound cannot be detected. \textcite{gordon_perceptual_1987} studied
perceptual attack times in different pairs of instruments. These have
variations that range from 6 to 25 ms, but in the flute they are about
10 ms. Minimum durations needed to identify pitch are typically 4
periods (less than 10 ms for flute notes)
\citep{patterson_threshold_1983}.

\section{Conclusions}
\label{sec:conclusions}

For single key transitions, the transition time is simply determined
by the time for a finger to push a key in one direction, typically 10
ms, or of a spring to push it in the other, typically 16 ms. When more
than one finger is involved, the delay times between individual key
movements must be added to this. For a transition involving only two
fingers and thus only two pathways, players in general coordinate
their fingers so that the unsafe transition is avoided. For some
transitions, there is no safe path. Professionals, unsurprisingly, are
more nearly simultaneous than amateurs. For both amateurs and
professionals, total delay increases with the number of fingers in
contrary motion.

\begin{acknowledgments}
  We thank our experimental subjects, The WoodWind Group and the
  Australian Research Council for their support.
\end{acknowledgments}

\end{document}